\documentclass[%
reprint,
nofootinbib,
superscriptaddress,
 amsmath,amssymb,
 aps,
 pra,
]{revtex4-1}

\usepackage{import}

\usepackage{graphicx}
\usepackage{physics}
\usepackage{dcolumn}
\usepackage{bm}

\usepackage[utf8]{inputenc} 

\usepackage{multirow}
\usepackage[warn]{mathtext}
\usepackage{amssymb}

\usepackage{caption}
\usepackage{subcaption}

\usepackage{textcomp} 
\usepackage{indentfirst} 
\usepackage{amsmath} 
\usepackage{graphicx}

\DeclareGraphicsExtensions{.pdf,.png,.jpg}
\usepackage{pgfplots}
\pgfplotsset{compat=1.13}

\usepackage{hyperref}
\hypersetup{
    colorlinks=true,
    linkcolor=blue,
    filecolor=black,      
    urlcolor=black,
    citecolor= violet
}

\usepackage{algpseudocode}

\usepackage{adjustbox}
\usepackage{tabularx}

\renewcommand{\bibname}{References}

\usepackage[nottoc]{tocbibind}
\usepackage{mathtools}

\usepackage{xcolor}
\definecolor{C0}{HTML}{4C72B0}
\definecolor{C1}{HTML}{DD8452}
\definecolor{C2}{HTML}{55A868}
\definecolor{C3}{HTML}{C44E52}



\newcolumntype{Y}{>{\centering\arraybackslash}X}

\makeatletter
\def\@fnsymbol#1{\ensuremath{\ifcase#1\or \dagger\or \ddagger\or
   \mathsection\or \mathparagraph\or \|\or **\or \dagger\dagger
   \or \ddagger\ddagger \else\@ctrerr\fi}}
\makeatother

\begin{document}

\title{Impact of qubit anharmonicity on near-resonant Rabi oscillations}

\author{Grigoriy~S.~Mazhorin}
\email{mazhorin.gs@phystech.edu}
\affiliation{National University of Science and Technology ``MISIS'', 119049 Moscow, Russia}
\affiliation{Russian Quantum Center, 143025 Skolkovo, Moscow, Russia}
\affiliation{Moscow Institute of Physics and Technology, 141701 Dolgoprudny, Russia}

\author{Andrei~A.~Kugut}
\affiliation{National University of Science and Technology ``MISIS'', 119049 Moscow, Russia}
\affiliation{Russian Quantum Center, 143025 Skolkovo, Moscow, Russia}
\affiliation{Moscow Institute of Physics and Technology, 141701 Dolgoprudny, Russia}

\author{Artyom~M.~Polyanskiy}
\affiliation{National University of Science and Technology ``MISIS'', 119049 Moscow, Russia}
\affiliation{Russian Quantum Center, 143025 Skolkovo, Moscow, Russia}
\affiliation{Moscow Institute of Physics and Technology, 141701 Dolgoprudny, Russia}

\author{Ilya~N.~Moskalenko}
\affiliation{National University of Science and Technology ``MISIS'', 119049 Moscow, Russia}

\author{Ilya~A.~Simakov}
\affiliation{National University of Science and Technology ``MISIS'', 119049 Moscow, Russia}
\affiliation{Russian Quantum Center, 143025 Skolkovo, Moscow, Russia}
\affiliation{Moscow Institute of Physics and Technology, 141701 Dolgoprudny, Russia}


\date{\today}

\begin{abstract}
Precise quantum control relies on a deep understanding of the dynamics of quantum systems under external drives. This study investigates the impact of anharmonicity on qubit dynamics under conditions typical for two-qubit entangling gates activated by weak near-resonant microwave drives. We measure the Rabi oscillation frequency as a function of drive amplitude and detuning. Our results reveal a linear dependence of the squared Rabi frequency on the squared drive amplitude, which relates to the ratio of detuning to anharmonicity, demonstrating strong agreement between experimental data and analytical predictions. Additionally, we analyze the leakage and phase errors arising from inaccurate Rabi frequency adjustments in the CZ gate implementation on fluxonium qubits driven by a microwave signal applied to the coupler~\cite{simakov2023coupler}.
\end{abstract}

\maketitle

\section{Introduction}
Quantum state manipulation requires accurate description of the fundamental phenomena governing the dynamics of driven quantum systems.
Factors such as the Stark shift, counter-rotating terms, and the multi-level structure of superconducting qubits impose constraints on manipulation techniques, while simultaneously facilitating the development of novel, more efficient methods of qubit manipulation.
Advanced quantum control, combined with substantial improvements in fabrication techniques, has resulted in remarkable achievements, including single-qubit gate fidelities exceeding 0.99997 \cite{rower2024suppressing} and 24-day stable two-qubit gates with fidelity surpassing 0.999 \cite{lin202424}.

One of the most prominent examples of the ambiguous impact on quantum computing is the multilevel structure of superconducting qubits. On one hand, additional levels introduce extra sources of errors, such as leakage and undesired Stark shifts, which need to be addressed by high-fidelity quantum control techniques \cite{werninghaus2021leakage,hyyppa2024reducing, steffen2003accurate}. On the other hand, these levels represent a hidden resource that can be utilized for improved initialization \cite{wang2024efficient}, readout \cite{elder2020high, zhu2013circuit}, two-qubit gate implementation \cite{ficheux2021fast, simakov2023coupler, ding2023high, moskalenko2022high}, and acceleration of quantum algorithms \cite{goss2024extending, PhysRevX.11.021010, PhysRevX.13.021028, PhysRevApplied.19.064024, nikolaeva2024scalableimprovementgeneralizedtoffoli}.

The effects of the multilevel structure of superconducting qubits are generally more pronounced under high-power drive conditions. Furthermore, the achievement of more precise, rapid and efficient quantum control often necessitates the application of strong drive fields. As a result, many studies have concentrated on this particular scenario \cite{schneider2018local, dutta2008multilevel, strauch2007strong, schon2020rabi}, yielding substantial experimental observations, including phenomena such as Autler-Townes shifts \cite{baur2009measurement, peng2018vacuum, FedorovGleb_2020}, Mollow triplets \cite{baur2009measurement}, multiphoton transitions \cite{shevchenko2012multiphoton, munyaev2021control}, electromagnetically induced transparency \cite{sun2014electromagnetically}, and multipartite entanglement \cite{lu2022multipartite}. However, the development of microwave-activated two- or many-qubit entanglement gates \cite{ficheux2021fast, ding2023high, simakov2024high} has introduced new drive conditions: weak near-resonant drive.

In this work, we study the behavior of the fluxonium qubit \cite{manucharyan2009fluxonium, moskalenko20219planar} under a weak near-resonant drive. We measure the Rabi oscillation frequency for the $0-1$ fluxonium transition as a function of drive amplitude and detuning for both of its sweet spots. We demonstrate that the coefficient relating the squared Rabi frequency to the squared drive amplitude contains a small-order correction term that linearly depends on the ratio of the signal detuning to the qubit anharmonicity. Additionally, we provide a theoretical analysis, which shows good agreement with experimental data. Although the multilevel structure’s impact on the $0-1$ transition Rabi frequency has been addressed in the literature for both weak \cite{strauch2007strong} and high anharmonicity \cite{schuster2005ac}, these studies primarily focus on strong resonant signals. The key difference in our work is the explicit consideration of detuned signals. Furthermore, we emphasize that ignoring the described effect  can diminish the fidelity of two-qubit CZ gates activated by a microwave pulse on the coupler \cite{simakov2023coupler}.
We evaluate the leakage and phase errors that arise from disregarding the considered contribution to the Rabi frequency.

\section{Model}
\begin{figure}[t]
    \center{\includegraphics[width=\linewidth]{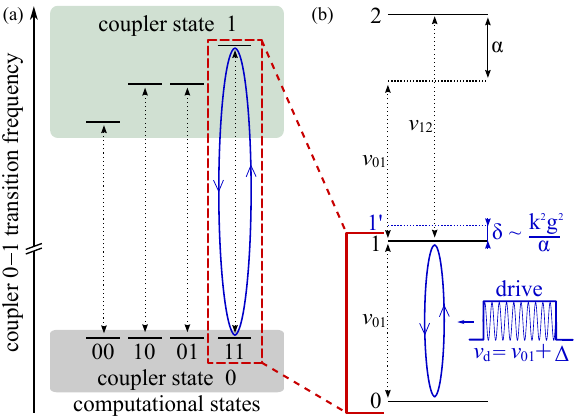}}
    \caption{(a) Schematic diagram of a two-qubit coupler microwave-activated controlled-phase gate. The main transition of the coupler splits into four distinct frequencies, each corresponding to a computational state. The near-resonant drive induces population oscillations only for the target $|11\rangle$ computational state.
    (b) Schematic diagram of the simplified three-level system. The drive interaction with the second excited coupler state results in a Stark shift $\delta$ of the main coupler transition $\nu_{01}$.}
    \label{fig:Single_qubit}
\end{figure}

We study Rabi oscillations in a superconducting qubit system under conditions representative of a two-qubit controlled-phase gate driven by a microwave pulse applied to a coupling element \cite{simakov2023coupler}. The principle of this operation is as follows: The frequency of the main coupler transition $0 - 1$ strongly depends on the states of the computational qubits. As a result, a direct $2\pi$-pulse applied to the coupler $0 - 1$ transition, associated with a target computational state (e.g., $|11\rangle$), leads to the effective accumulation of a phase $\pi$ on this state. Figure~\ref{fig:Single_qubit}(a) schematically illustrates the energy shift of the coupler main transition for discrete computational states, as well as the gate concept.

A notable feature of this method is the near-resonant drive applied to the coupler. 
This requirement arises from the fact that even a weak external signal tuned to the target transition frequency, while having no significant effect on the coupler's ground state associated with the other computational states, can still induce small, unwanted phase accumulations in those states \cite{simakov2024high}.
A simple way to compensate this effect is to slightly detune the drive signal from resonance, ensuring that the total phase accumulated across all computational states results in the desired $\pi$ phase.

The two-qubit operation can be effectively reduced to a single-qubit gate applied directly to the coupler. Similar to a single-qubit gate, to suppress unwanted leakage to high-energy levels, the drive signal should be sufficiently weak compared to the coupler anharmonicity. Additionally, it is generally preferable for the signal to be weaker than the dispersive shift of the coupler induced by its strong interaction with the computational qubits, although this is not a strict requirement. Furthermore, since the coupler $0 - 1$ transition frequency significantly exceeds the signal detuning, it is appropriate to apply the rotating wave approximation (RWA) to analyze the system.

Additionally, we emphasize an important advantage of the microwave-activated two-qubit gate: Throughout the operation, both the qubits and the coupler are maintained at their sweet spots, where they exhibit maximal coherence. This configuration also yields symmetric wavefunctions, thereby suppressing transitions of equal parity, such as $0 - 2$.

Ultimately, we summarize the above considerations into five assumptions, which form the basis of our analytical analysis: (i) The system can be effectively described by a three-level model. (ii) The RWA is valid. (iii) The drive is near-resonant, but the detuning $\Delta$ is small compared to the anharmonicity $\alpha$: $\Delta \ll \alpha$. (iv) The drive strength $g$ is weak, satisfying $g \ll \alpha$. (v) The equal-parity transition $0-2$ is forbidden. As a result, the multi-qubit problem is reduced to a single-qubit system (see Figure~\ref{fig:Single_qubit}(b)), which can be solved analytically and experimentally investigated using just a single-qubit device.

Under introduced assumptions, the Hamiltonian takes the following form:
\begin{equation}
    \renewcommand{\arraystretch}{1.7}
    \hat{H}_{\mathrm{RWA}} = \begin{pmatrix}
  \textcolor{white}{00}0\textcolor{white}{00}& \textcolor{white}{00}\displaystyle\frac{g}{2}\textcolor{white}{0} & 0\\
  \displaystyle\frac{g}{2}& -\Delta & k\displaystyle\frac{g}{2}\\
  0& k\displaystyle\frac{g}{2}& -2\Delta+\alpha
\end{pmatrix},
    \label{Ham}
\end{equation}
where $g$ is the drive amplitude; $\Delta = \nu_d - \nu_{01}$ is the detuning and $\alpha=\nu_{12}-\nu_{01}$ is the system anharmonicity; $\nu_{d}$, $\nu_{01}$, $\nu_{12}$ are drive, transition $0-1$ and transition $1-2$ frequencies; $k = m_{12}/m_{01}$; $m_{01}$ and $m_{12}$ are the matrix elements corresponding to transitions $0-1$ and $1-2$ respectively.

Considering the condition $g \ll \alpha$, we compute the eigenvalues of the Hamiltonian and derive the following expression for the $0-1$ transition Rabi frequency (see Appendix~\ref{appendix:analytic} for details):

\begin{equation}
    \begin{split}
        \Omega_{\mathrm{Rabi}}^2 &= \displaystyle\alpha^2 \, \left( \frac{\Delta^2}{\alpha^2} + \frac{g^2}{\alpha^2} + \frac{k^2}{2}\frac{\Delta\,g^2}{\alpha^3} + \overline{\mathrm{O}}\left(\frac{g^4}{\alpha^4}\right) \right) \\
    &\approx \displaystyle\Delta^2 + g^2\left(1+\frac{k^2}{2}\frac{\Delta}{\alpha}\right).
    \end{split}
    \label{rabi_correction}
\end{equation}

We note that the coefficient linking the squared Rabi frequency to the squared drive amplitude contains a small correction term that linearly depends on the ratio between the signal detuning and the qubit anharmonicity. We denote this coefficient as $s$:
\begin{equation}
s = 1+\frac{k^2}{2}\frac{\Delta}{\alpha}.
\label{slope}
\end{equation}

This linear correction, as a physical phenomenon, is attributed to the Stark shift $\delta$, which arises from the interaction of the drive with the $1-2$ transition:
\begin{equation}
\delta = \frac{k^2}{4}\frac{g^2}{\alpha}.
\label{delta_freq}
\end{equation}

\begin{figure*}[t]
    \center{\includegraphics[width=\linewidth]{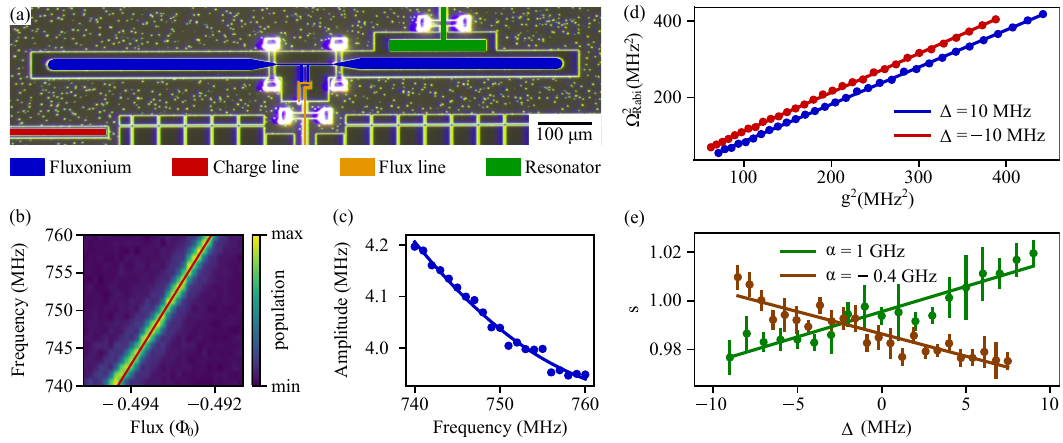}}
    \caption{The device and the experimental results. (a)~False-colored optical micrograph of the fluxonuim qubit~(blue) along with its readout resonator~(green), charge line~(red) and local flux line~(orange). (b)~Rabi spectroscopy measurement. Red line shows the qubit frequency dependence on the external magnetic flux. (c)~Frequency-dependent transfer function of the signal line. (d)~The squared Rabi oscillation frequency $\Omega^2_{\mathrm{Rabi}}$ dependence on the squared near-resonant signal amplitude $g^2$ for different detunings~$\Delta$. The points depict the experimental data and the solid lines show the linear fits.  (e)~The slope~$s$ dependence on the signal detuning~$\Delta$ for different qubit anharmonicities~$\alpha$. The points correspond to the experimental data and the solid lines show the linear fits. The data for different signs of qubit anharmonicity have different trends according to the equation~(\ref{slope}).} 
    \label{fig:experiment}
\end{figure*}

The Rabi frequency characterizes the speed of the coupler population oscillations and correspondingly the rate of phase accumulation. Consequently, this parameter directly determines the most significant coherent errors associated with the two-qubit entangling gate described above, namely the residual population and accumulated phase inaccuracies. We believe that our straightforward analysis will aid identifying optimal design parameters and improving calibration techniques for implementing high-fidelity microwave-activated controlled phase gates. Further, we analyze the impact of Rabi frequency adjustments on the coherent gate errors.

\section{Experiment}

We experimentally investigate the impact of anharmonicity on the Rabi frequency of the $0-1$ transition using a fluxonium qubit. The colored optical image of the single-qubit device is given in Fig.~\ref{fig:experiment}(a). It consists of a floating fluxonium qubit (blue) that is capacitively coupled to a $\lambda$/4 readout resonator (green) and a charge line (red), and galvanically connected to a flux line (orange). Besides the fluxonium mode, the qubit also has an additional harmonic mode \cite{moskalenko2021tunable}. In our design, the harmonic mode frequency is engineered to differ by more than 1~GHz from the sweet spot frequencies of the fluxonium qubit, ensuring that it does not affect the qubit dynamics.
The anharmonicity $\alpha$ and the ratio $k$ are calculated using fluxonium parameters obtained from the spectroscopy data analysis (see Appendix~\ref{appendix:parameters}). The pulse measurements are conducted near the low and high sweet spots, where the qubit frequency $\nu_{01}$ is $ 0.75$~and~$4.66$~GHz, the anharmonicity $\alpha$ is equal to $1.335$~and~$-0.403$~GHz, and the ratio $k$ is $2.44$~and~$1.35$, respectively.

We use the charge line for qubit excitation near the low sweet spot and the flux line near the high sweet spot. This choice is motivated by the goal to amplify the anharmonicity impact on the $0-1$ Rabi oscillations, achieved by increasing the ratio $k = m_{12}/m_{01}$, where $m_{ij}$ can be elements of either the charge $n_{ij}$ or flux $\varphi_{ij}$ matrices.
Since $n$ and $\varphi$ are conjugate variables $n_{ij}/\varphi_{ij} \sim \nu_{ij}$, one gets:
\begin{equation} 
    \frac{n_{12}}{n_{01}} \sim \frac{\nu_{12}}{\nu_{01}} \cdot \frac{\varphi_{12}}{\varphi_{01}}. 
\end{equation}

Given that the ratio $\nu_{12}/\nu_{01}$ is approximately $2.7$ at the low sweet spot and $0.9$ at the high sweet spot, we increase the ratio $k$ by using the charge line for the external drive at the low sweet spot and the flux line at the high sweet spot.

Since we aim to observe a low-magnitude effect of the anharmonicity impact on the $0-1$ Rabi oscillations, we need to exclude the influence of other potential sources.
One such source is the frequency-dependent transfer function of the signal line. The amplitude of the signal reaching the chip differs significantly from that emitted by the control equipment due to additional components, such as cables, attenuators, mixers, and filters. Impedance mismatching, the skin effect in cables \cite{4056481}, and the characteristics of the powder filter lead to a complex frequency dependence of the signal amplitude. To account for this effect in the subsequent data processing, we first measure the transfer function of the signal line, using the qubit as a signal detector.

A preliminary step involves performing Rabi spectroscopy. We apply a weak rectangular $\pi$-pulse, calibrated at the sweet spot, with a drive amplitude chosen to be sufficiently small such that the resulting Stark shift is negligible. Next, we measure the final population of the first excited state as a function of pulse detuning and external magnetic flux, both with and without the applied $\pi$-pulse. The latter case is necessary to mitigate the influence of changes in the resonator frequency on the readout state. The difference between these two cases, as a function of signal frequency and external magnetic flux, is shown in Fig.~\ref{fig:experiment}(b). The solid red line represents the qubit frequency dependence on the external magnetic flux. This procedure yields a calibration curve for the qubit frequency tuning.

Eventually, we are ready to measure the frequency-dependent transfer function of the signal line. Utilising the obtained calibration curve, we tune the qubit along the frequency range of interest, applying a resonant signal of the same amplitude at each point, and measure the frequency of the Rabi oscillations. Since the drive is always resonant, we directly observe the frequency-dependent transfer function of the signal line.
A typical result is shown in Fig.~\ref{fig:experiment}(c). The data are then smoothed using a Savitzky-Golay filter (solid line). In the subsequent experiments, we use these dependencies to map the incoming signal to the output signal from the arbitrary waveform generator.

To verify the equation~(\ref{rabi_correction}), we measure near-resonant Rabi oscillations
as a function of signal amplitude and detuning. The experiment is conducted near both fluxonium sweet spots. The dependence of the squared Rabi frequency on the squared drive amplitude for two different detunings is shown in Fig.~\ref{fig:experiment}(d).
The data are well-approximated by a linear function, and, according to the theoretical prediction~(\ref{rabi_correction}), the line gradients (\ref{slope}) are visibly different since the detunings differ by 20 MHz.

We also collect similar data for a varying detuning near both sweet spots. The resulting slope dependencies of the detuning for the two sweet spots are depicted in Fig.~\ref{fig:experiment}(e). The trends in these dependencies are opposite, reflecting the sign of the anharmonicity. We fit the dependencies using linear functions, yielding coefficients of $2.08\pm0.18$ ns for the low sweet spot and $-2.16\pm0.19$ ns for the high sweet spot. These values are in good agreement with the theoretical predictions of $2.23$ ns and $-2.26$ ns, respectively.

\section{Discussion}

Although the qubit anharmonicity has a subtle effect on the Rabi oscillation frequency of a single qubit, its impact becomes more significant for gate performance when the qubit is part of a larger circuit and its quantum state manipulation is used for multi-qubit entanglement. To illustrate this, we consider the approach for implementing a two-qubit microwave gate on fluxonium qubits~\cite{simakov2023coupler}.

\begin{figure}[t]
    \center{\includegraphics[width=\linewidth]{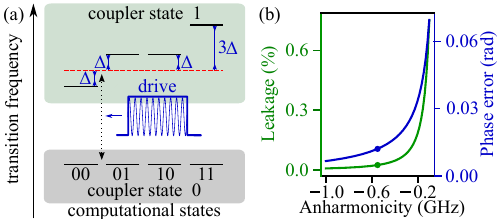}}
    \caption{(a) State-dependent spectrum of the coupler $0-1$ transition. Applied drive induces near-resonant coupler Rabi oscillations for all computational states. The drive frequency (red dashed line) is positioned in the midpoint between the frequencies of the main coupler transitions for the non-excited and single-excited states of the two-qubit system. (b) The leakage and phase error dependence on the coupler anharmonicity.}
    \label{fig:Two_qubit}
\end{figure}

The gate concept is illustrated in Fig.~\ref{fig:Two_qubit} (a). The drive applied to the coupler, with amplitude $g = \sqrt{\frac{5}{3}} \Delta$, frequency $\nu_d = (\nu_{00} + \nu_{10}) / 2$, and duration $\tau = \sqrt{\frac{3}{2}} \pi / \Delta$, results in an ideal CZ gate when considering only the interaction with the $0-1$ coupler transition. We evaluate the leakage and phase error caused by the Stark shift of the coupler frequency using the following parameters: signal detuning $\Delta = 14$ MHz, as shown in Fig.~\ref{fig:Two_qubit}(a), anharmonicity $\alpha=-550$ MHz and the ratio of the matrix elements for the $1-2$ to $0-1$ transitions $k=1.29$, which is similar to the experimentally obtained parameters of the device presented in Ref.~\cite{simakov2023coupler}. To do this, we calculate the amplitude $a_0^{mn}$  of the ground coupler state after the applied drive for each computational state $mn~\in~\{|00\rangle, |01\rangle, |10\rangle, |11\rangle \}$ using the following equation:
\begin{equation}
    a_0^{mn} = (i\Delta^{mn}/\Omega^{mn}_{\mathrm{Rabi}})\sin(\Omega^{mn}_\mathrm{Rabi}t/2)+\cos(\Omega^{mn}_\mathrm{Rabi}t/2),
\end{equation}
where $\Delta^{mn}$ is the frequency detuning from the drive to the corresponding $0-1$ coupler transition, and $\Omega^{mn}_{\mathrm{Rabi}}$ is the Rabi frequency, calculated using the formula in Eq.~(\ref{rabi_correction}).

The resulting leakage and phase error are $0.02\%$ and $0.016$ radians, respectively. Furthermore, we calculate these errors as a function of coupler anharmonicity, with the results presented in Fig.~\ref{fig:Two_qubit} (b). These errors can be entirely eliminated through frequency calibration that accounts for shifts caused by the multilevel structure.
Therefore, to achieve a high-fidelity two-qubit gate, it is crucial to account for the drive interaction with the second state, especially for couplers with low anharmonicity.

\section{Conclusion and outlook}

In this work, we investigate the Rabi oscillations in superconducting qubit systems induced by a weak, near-resonant drive. Leveraging the rich energy spectrum of a fluxonium qubit, we experimentally demonstrate that the ratio between the squared Rabi frequency and the squared drive amplitude depends on both the signal detuning and the qubit anharmonicity. Furthermore, we model the system using an effective three-level approximation and analytically derive a small-order correction to the coefficient relating the squared Rabi frequency to the squared drive amplitude, which is proportional to the ratio of the drive detuning to the anharmonicity. We find good agreement between the theoretical and experimental results over the studied parameter range and explain the phenomenon by a Stark shift of the $0-1$ transition frequency induced by the interaction between the drive and the $1-2$ qubit transition.

From the perspective of quantum computing, the observed effect needs to be taken into account
during the calibration procedure of multi-qubit gates activated by a microwave drive on the coupler.
We analyze the corresponding leakage and phase errors induced by small adjustment to the Rabi frequency in the implementation of the CZ gate. The impact on the gate fidelity increases as the anharmonicity decreases. Therefore, it is important to account for this effect in the case of couplers with low anharmonicity.

\section*{Acknowledgments}

The authors are grateful to Ilya Besedin and Alexey Ustinov for valuable discussions and feedback on the manuscript. 
We also thank Daria Kalacheva, Victor Lubsanov and Alexey Bolgar for fabricating the sample.
We acknowledge partial support from the Ministry of Science and Higher Education of the Russian Federation in the framework of the Program of Strategic Academic Leadership “Priority 2030” (MISIS Strategic Project Quantum Internet). 

\appendix
\section{Exact solution analysis}
\label{appendix:analytic}
Here, we provide an exact analytical solution for the Rabi oscillation frequency. By solving the cubic characteristic equation of the Hamiltonian (\ref{Ham}), we find its eigenvalues:
\begin{equation}
    \lambda_{r} = 2\,\sqrt{\frac{I_1}{3}}\,\cos\left(\frac{1}{3}\,\arccos{\frac{I_2}{2 I_1^{3/2}}-\frac{2\pi r}{3}}\right),
    \label{precise_rabi}
\end{equation}
where
\begin{equation*}
    \renewcommand{\arraystretch}{3}
    \begin{array}{cccc}
    I_1 = \displaystyle\frac{3}{4} g^2(1+k^2)+\alpha^2-3 \alpha \Delta+3 \Delta^2,
    \\
    I_2 = \displaystyle-\frac{9 g^2 \alpha}{2}+\frac{9}{4} g^2 k^2 \alpha+2 \alpha^3+\frac{27 g^2 \Delta}{4}
    \\
    \displaystyle\quad\quad\quad\quad\,\,-\frac{27}{4} g^2 k^2 \Delta-9 \alpha^2 \Delta+9 \alpha \Delta^2,
    \\
    \displaystyle r = 0,1,2.
    \end{array}
\end{equation*}

The corresponding Rabi frequencies, determined as $|\lambda_i-\lambda_j|$, take the following form:

\begin{equation}
    \Omega_{i} = 2\,\sqrt{\frac{I_1}{3}}\,\left|\,\sin \left(\varphi_i - \frac{1}{3}\,\arcsin{\frac{I_2}{2 I_1^{3/2}}}\right)\right|,
    \label{precise_rabi}
\end{equation}
where
\begin{equation*}
    \displaystyle\varphi_0 = \frac{\pi}{2},\; \varphi_1 = \frac{\pi}{6},\; \varphi_2 = -\frac{\pi}{6}.
\end{equation*}

To simplify the Rabi frequency, we utilize a well-known trigonometric identity:
\begin{equation*}
    \sin{3\gamma} = 3 \, \sin{\gamma} - 4 \, \sin^{3} \gamma,
\end{equation*}
making a substitution of variables:
\begin{equation*}
    \gamma = \varphi_i - \frac{1}{3}\arcsin{\frac{I_2}{2 I_1^{3/2}}},
\end{equation*}
we get the equation on Rabi frequencies.
To eliminate the influence of the signs, we square the result, yielding
a cubic equation for $\Omega^2$ that encompasses all $\{\Omega_i\}$:
\begin{equation}
    4\, I_1^3 - I_2^2 = 27\,\Omega^2\,\left(I_1 - \Omega^2\right)^2.
    \label{ansatz}
\end{equation}
We find the root, corresponding to the 0-1 transition in the range $|\alpha-\frac{3}{2}\Delta|>\sqrt{\Delta^2+g^2}$:
\begin{equation}
    \Omega_{\mathrm{Rabi}}^2 = \frac{2}{3}I_1 - \frac{1}{3}\left( I_0^{1/3} + \frac{I_1^2}{I_0^{1/3}} \right),
    \label{final_rabi}
\end{equation}
where
\begin{equation*}
    I_0 = - I_1^3 + \frac{I_2^2}{2} + \frac{I_2}{2} \, \sqrt{I_2^2 - 4 I_1^3}.
\end{equation*}
Finally assuming $\Delta \sim g \ll \alpha$ we can get the following approximation of (\ref{final_rabi}):
\begin{equation}
    \renewcommand{\arraystretch}{3}
    \begin{array}{ll}
    \Omega_{\mathrm{Rabi}}^2 = \displaystyle\alpha^2 \, \left( \frac{\Delta^2}{\alpha^2} + \frac{g^2}{\alpha^2}\,\left(1 + \frac{k^2}{2}\frac{\Delta}{\alpha}\right) + \overline{\mathrm{O}}\left(\frac{g^4}{\alpha^4}\right) \right)
    \\
    \quad\quad\,\,\,\,\approx \displaystyle\Delta^2 + g^2\left(1+\frac{k^2}{2}\frac{\Delta}{\alpha}\right).
    \end{array}
\end{equation}

\begin{figure*}
    \center{\includegraphics[width=\linewidth]{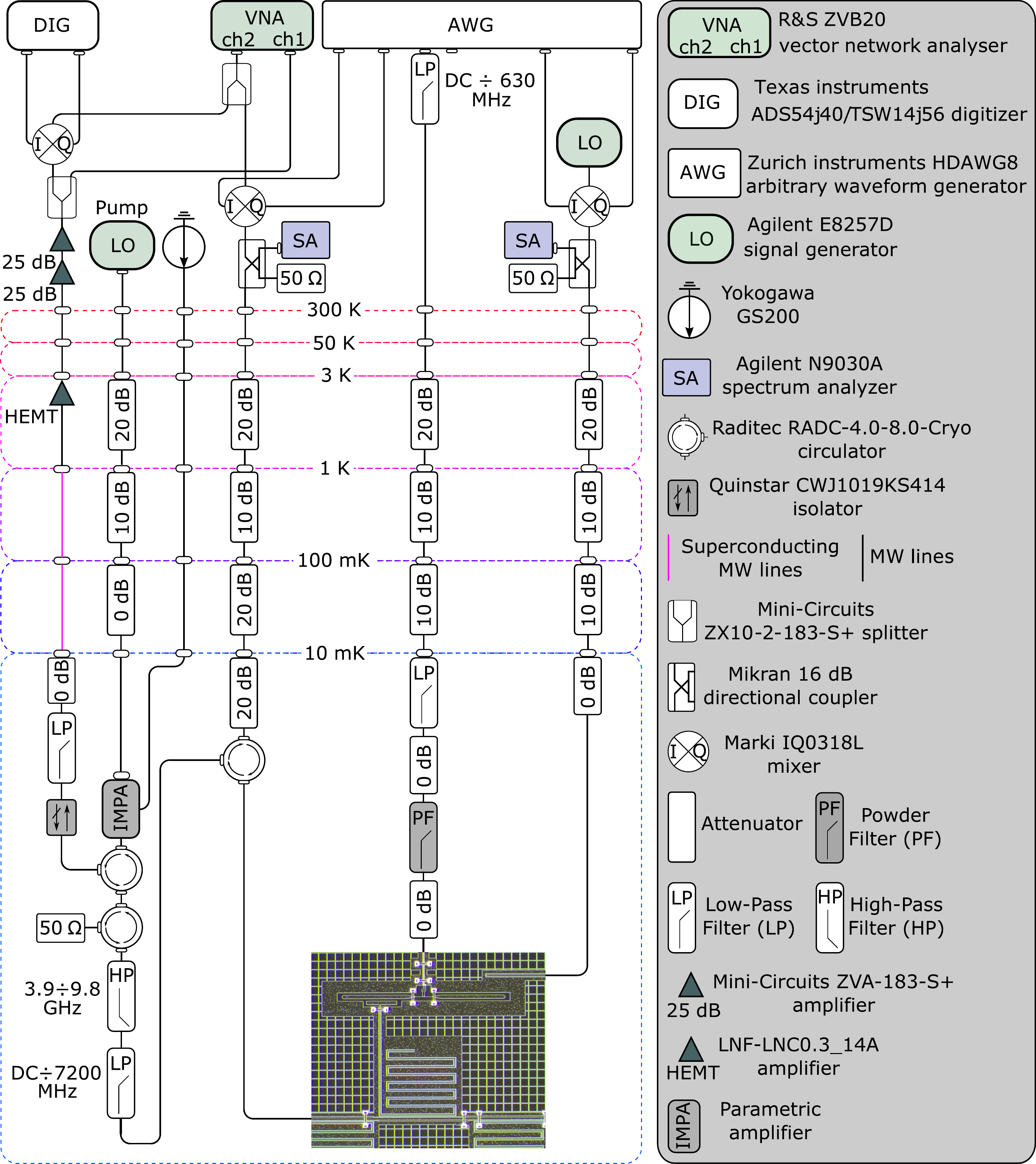}}
    \caption{Schematic diagram of the experimental setup.}
    \label{fig:Scheme_equip}
\end{figure*}

\section{Experimental setup}
\label{appendix:exp_setup}

The experimental setup is depicted in Fig.~\ref{fig:Scheme_equip}. Experiments are conducted using a BlueFors LD-250 dilution refrigerator, which maintains a base temperature of 10 mK. The chip is interfaced with the control system via three lines: the readout line, the flux control line (Z control), and the excitation charge control line (X control). Pulse generation and flux control are performed by a Zurich Instruments HDAWG8 arbitrary waveform generator.

IQ microwave mixers are utilized to up-convert and down-convert the intermediate frequency readout pulses to and from the resonator frequencies. After reflecting off the qubit chip, the readout microwave signal is analyzed using a vector network analyzer (R\&S ZVB20) for spectroscopy, and a custom-built digitizer setup for single-shot readout. Mixer calibration is performed with a spectrum analyzer (Agilent N9030A).

We use one analog port to control the flux in the fluxonium circuit. Due to the significant difference between the fluxonium frequencies at the high and low sweet spots, we excite the qubit either directly from the generator or via an additional IQ mixer for frequency conversion. In Fig.~\ref{fig:Scheme_equip} we show only the second configuration.

Microwave attenuators are used to mitigate the influence of thermal and instrumental noise from signal sources at room temperature on the qubit chip. The readout line includes an impedance-matched parametric amplifier (IMPA), followed by a Quinstar CWJ1019KS414 isolator to prevent noise from higher temperature stages from affecting the IMPA and the qubit device. The IMPA is pumped using an Agilent E8257D signal generator. Three Raditec RADC-4.0-8.0-Cryo circulators, along with a set of low-pass and high-pass filters placed after the sample, protect it from IMPA pumping and reflected signals.

At the PT2 stage (3 K) of the cryostat, an LNF-LNC0.3 14A high-electron-mobility transistor (HEMT) is installed. The output line is further amplified outside the cryostat using two Mini-Circuits ZVA-183-S+ amplifiers. A low-pass filter (Mini Circuits VLF-630+) is used in conjunction with a powder filter, which provides 15 dB attenuation near the qubit frequency in the flux control line, while no filter is used in the excitation charge control line.

\section{Single qubit parameters}
\label{appendix:parameters}
In our experiments, we use two fluxonium qubits, labeled A and B, which are designed to be identical. The fluxonium obeys the Hamiltonian:

\begin{equation}
    \frac{\hat{H}}{h} = 4E_C\hat{n}^2 + \frac{E_L}{2}\hat{\varphi}^2 - E_J \cos\left(\hat{\varphi} - \frac{2\pi \Phi_e}{\Phi_0}\right),
\end{equation}
where \( E_C \), \( E_L \), and \( E_J \) are the charge, inductive and Josephson energies of the fluxonium, respectively. The operators \( \hat{\varphi} \) and \( \hat{n} \) represent the phase and charge operators, and \( \Phi_e \) and \( \Phi_0 \) denote the external and quantum magnetic fluxes, respectively.

We perform two-tone spectroscopy as a function of the external magnetic flux in order to obtain the device parameters. The extracted parameters are summarized in Table~\ref{table:parameters}. These derived parameters are then used to calculate the fluxonium characteristics presented in the main text. The results near the low sweet spot are obtained using fluxonium A, while the results near the high sweet spot are obtained using fluxonium B.

\begin{table}[t]
\centering
\begin{tabularx}{\columnwidth} { 
     >{\raggedright\arraybackslash}X 
     >{\centering\arraybackslash}X 
     >{\centering\arraybackslash}X 
     >{\centering\arraybackslash}X}
\hline
\hline
\textbf{Fluxonium} & \( E_C \)/\(h\), GHz & \( E_L \)/\(h\), GHz & \( E_J \)/\(h\), GHz \\
\hline
\textbf{A} & 0.49 & 1.74 & 3.56 \\
\textbf{B} & 0.5 & 2.11 & 4.29\\
\hline
\hline
\end{tabularx}
\caption{Device parameters for fluxoniums A and B.}
\label{table:parameters}
\end{table}

\renewcommand{\bibname}{Reference}
\bibliographystyle{unsrtnat}
\bibliography{main}

\end{document}